\documentclass[letterpaper, 10 pt, conference]{ieeeconf}  

\IEEEoverridecommandlockouts                              
\usepackage{graphicx, , amsmath}
\usepackage{amsfonts, amssymb}
\IEEEoverridecommandlockouts
\usepackage{mdwmath}
\usepackage{texdraw}%
\usepackage{psfrag}%
\usepackage{accents}%
\usepackage{color}%
\usepackage{cite}
\usepackage{graphics}
\usepackage[tight,footnotesize]{subfigure}

\newtheorem{remark}{Remark}

\title{\LARGE \bf
A Deterministic Annealing Optimization Approach for Witsenhausen's and Related Decentralized Control Settings
}

\author{Mustafa Mehmetoglu, Emrah Akyol, and Kenneth Rose
\thanks{This work was supported in part by the NSF under grant CCF-1118075}
\thanks{Mustafa Mehmetoglu and Kenneth Rose are with Department of Electrical and Computer Engineering, University of California at Santa Barbara, CA 93106, USA,
         {\tt\small \{mehmetoglu, rose\}@ece.ucsb.edu}}%
\thanks{Emrah Akyol is with the Department of Electrical Engineering, University of Southern California, Los Angeles, CA, 90089, USA,
{\tt\small{eakyol@usc.edu}}}%
}

\begin{document}

\maketitle
\thispagestyle{empty}
\pagestyle{empty}

\begin{abstract}
This paper studies the problem of mapping optimization in decentralized control problems. A global optimization algorithm is proposed based on the ideas of ``deterministic annealing" - a powerful non-convex optimization framework derived from information theoretic principles with analogies to statistical physics. The key idea is to randomize the mappings and control the Shannon entropy of the system during optimization. The entropy constraint is gradually relaxed in a deterministic annealing process while tracking the minimum, to obtain the ultimate deterministic mappings. Deterministic annealing has been successfully employed in several problems including clustering, vector quantization, regression, as well as the Witsenhausen's counterexample in our recent work\cite{mehmetoglu2014deterministic}. We extend our method to a more involved setting, a variation of Witsenhausen's counterexample, where there is a side channel between the two controllers. The problem can be viewed as a two stage cancellation problem. We demonstrate that there exist complex strategies that can exploit the side channel efficiently, obtaining significant gains over the best affine and known non-linear strategies.
\end{abstract}

\section{Introduction}
Decentralized control systems have multiple controllers designed to collaboratively achieve a common objective while taking actions based on their individual observations. No controller, in general, has direct access to the observations of the other controllers. This makes the design of optimal decentralized control systems a very challenging problem. One of the most studied structures, termen ``linear quadratic Gaussian" (LQG), involves linear dynamics, quadratic cost functions and Gaussian distributions. Since in the case of centralized LQG problems, the optimal mappings are linear, it was naturally conjectured that linear control mappings remain optimal even in decentralized settings. However, Witsenhausen proposed in \cite{witsenhausen} an example of a decentralized LQG control problem, commonly referred to as Witsenhausen's counterexample (WCE), for which he provided a simple non-linear control strategy that outperforms all affine strategies.

Decentralized control systems such as WCE arise in many practical applications, and numerous variations of WCE have been studied in the literature (see, e.g., \cite{basar2008variations,ubli}). One example introduced in \cite{martins} considers a two stage noise cancellation problem. This variant includes an additional noisy channel over which the two controllers can communicate. The second controller, therefore, has access to some (corrupted) side information which is controlled by the first controller. We refer to this setting as the ``side channel problem" motivated by the class of "decoder side information" problems in communications and information theory\cite{el2011network}. Specifically, this problem is a zero-delay source-channel coding variation of the coded side information problem studied in the seminal papers of Wyner \cite{wyner1975source}, and Ahlswede and Korner \cite{ahlswede1975source}.  It has been demonstrated in \cite{martins} that nonlinear strategies may outperform the best affine strategies, however, the question of how to approach the optimal solution remains open. Finding the optimal mappings for such problems is usually a difficult task unless they admit an explicit (and usually as simple as linear) solution, see e.g. \cite{basar2008variations} for a set of problems, some are tractable and others not.

In prior work\cite{mehmetoglu2014deterministic}, we proposed an optimization method, derived from information theoretic principles, which is suitable to a class of decentralized control problems. Specifically, the method was successfully employed for WCE and the best known cost for this benchmark problem was obtained. The method proposed in this work is an extension of our prior work, developed to account for the complex effects of the side channel problem introduced in \cite{martins}. The introduction of the side channel in this setting results in complex mappings that are highly nontrivial.

Deterministic annealing (DA) is motivated by statistical physics, but derived from basic principles in information theory. It has been successfully used in non-convex optimization problems, including clustering \cite{rose1990statistical}, vector quantization \cite{rose1992vector}, and more (see review in \cite{da}). DA introduces controlled randomization into the optimization process by incorporating a constraint on the level of randomness (measured by Shannon entropy) while minimizing the expected cost of the system. The resultant Lagrangian functional can be viewed as the ``free energy" of a corresponding physical system, and the Lagrangian parameter as the ``temperature". The optimization is equivalent to an annealing process that starts by minimizing the cost (free energy) at a high temperature, which effectively maximizes the entropy. The minimum cost is then tracked at successively lower temperatures as the system typically undergoes a sequence of phase transitions through which the complexity of the solution (mappings) grows. As the temperature approaches zero, hard (nonrandom) mappings are obtained.
  
In Section II we give the problem definition. In Section III we describe the proposed method, and in Section IV the experimental results are given. Discussion and concluding remarks are in Section V.

\section{Problem Definition}
Let $\mathbb E(\cdot)$, $\mathbb E\{\cdot|\cdot\}$  and  $\mathbb P(\cdot)$ denote the expectation, conditional expectation and probability operators, respectively. $H(\cdot)$ and $H(\cdot|\cdot)$ are the entropy and conditional entropy. $\mathbb R$ denotes the set of real numbers. The gaussian density with mean $\mu$ and variance $\sigma^2$ is denoted as ${\cal N} (\mu, \sigma^2)$.

\begin{figure}
\begin{minipage}[b]{1\linewidth}
  \centering
  \centerline{\includegraphics[width=8.5cm]{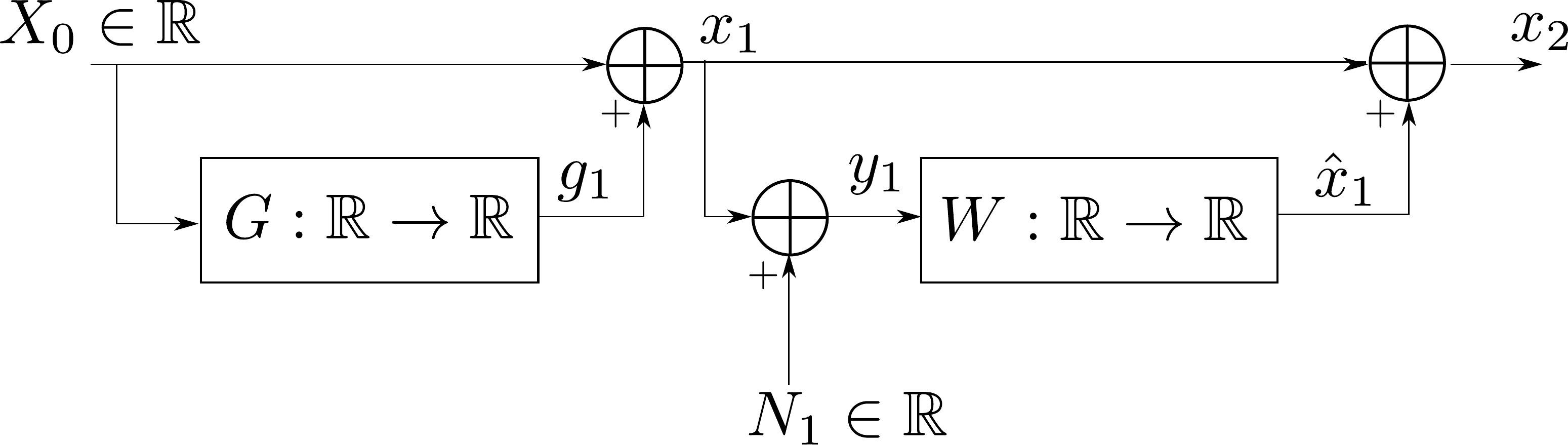}}
  \centerline{(a)}\medskip
\end{minipage}
\begin{minipage}[b]{1\linewidth}
  \centering
  \centerline{\includegraphics[width=8.5cm]{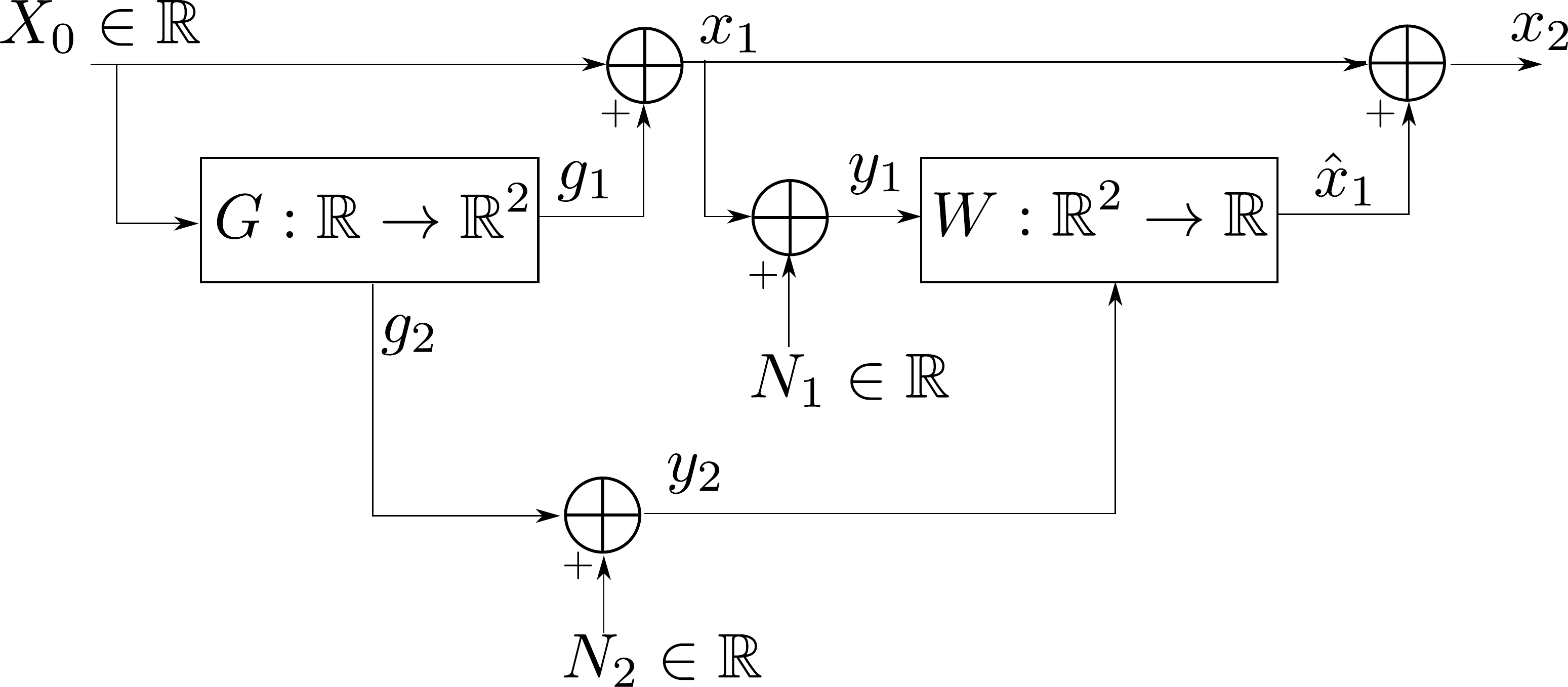}}
  \centerline{(b)}\medskip
\end{minipage}
\caption{Problem Settings for (a) original WCE and (b) side channel variation.}
\label{settings}
\end{figure}

\subsection{Original WCE}
The problem setting for the original WCE is given for reference purposes, and depicted in Figure \ref{settings}a. The source $x_0 \sim {\cal N} (0, \sigma_{x_0}^2)$ and noise $n \sim {\cal N} (0, 1)$ are independent. The two controllers ${ G}:\mathbb R\rightarrow \mathbb R$ and ${ W} :  \mathbb R  \rightarrow  \mathbb R$ aim to minimize the cost
\begin{equation}
D = \mathbb E \{k^2x_1^2+x_2^2\}
\label{eq:cost1}
\end{equation}
where $x_1= x_0+g(x_0)$ and $x_2=x_1-h(x_1+n)$. The given constant $k^2$ governs the trade-off between the control cost $\mathbb E \{x_1^2\}$ and the estimation error $\mathbb E \{x_2^2\}$.

\subsection{Side Channel Variation}
The following two-stage control problem was introduced in \cite{martins}:
\begin{align}
 x_1 &= x_0 + g_1 \\
 y_1 &= x_1 + n_1 \\
 x_2 &=x_1-\hat x_1
 \end{align}
where $x_0 \sim {\cal N} (0, \sigma_{X_0}^2)$ and $n_1 \sim {\cal N} (0, 1)$. The problem setting for this problem is given in Figure \ref{settings}b. There are two admissible controllers given by:
\begin{align}
\begin{bmatrix}g_1\\g_2\end{bmatrix}&=G(x_0) \\
\hat x_1&=W(y_1,z)
\end{align}
where $z = g_1+n_2$ and $n_2 \sim {\cal N} (0, \sigma_{n_2}^2) $. $x_0$, $n_1$ and $n_2$ are mutually independent. The problem is to find the optimal controllers $G: \mathbb R \rightarrow \mathbb R^2 $ and $W: \mathbb R^2 \rightarrow \mathbb R$ that minimize the cost 
\begin{equation}
D = k_1^2 \mathbb E\{(g_1)^2\} + k_2^2 \mathbb E\{g_2^2\} + \mathbb E\{x_2^2\}
\label{cost}
\end{equation}
for given $\sigma_{x_0}$, $\sigma_{n_2}$ and positive parameters $k_1$, $k_2$. The addition of the side channel over the original WCE problem is evident in Figure \ref{settings}.

The cost function defined in \cite{martins} does not include the term $\mathbb E\{g_2^2\}$. Instead, the cost is minimized subject to the following constraint:
\begin{equation}
\frac{\sqrt{\mathbb E\{g_2^2\}}}{\sigma_{n_2}}\le b_{SNR}
\end{equation}
for a given $b_{SNR}$. Side channel signal to noise ratio (SNR) is therefore $b_{SNR}^2$. We incorporate this constraint into the cost function by forming an overall Lagrangian cost with $k_2$ as Lagrange parameter. Different SNR values are obtained depending on the value of $k_2$.

\begin{figure}
\centerline{\includegraphics[width=8.5cm]{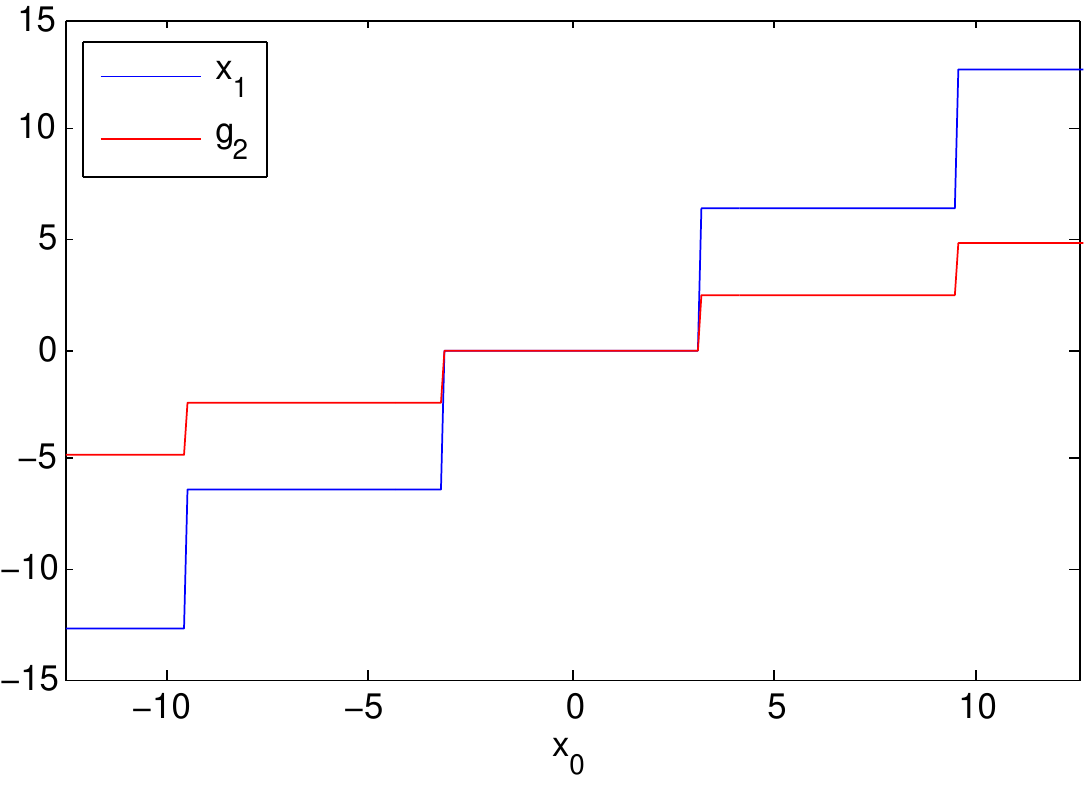}}
\caption {Mappings suggested in \cite{martins} for the side channel variation problem, where both mappings suggested are simple staircase functions. This example is for $b_{SNR} = 2$.}
\label{nuno_ex}
\end{figure}

The simple nonlinear mappings suggested in \cite{martins}, which widely outperform the best affine solution in a large range of SNR values, are depicted in Figure \ref{nuno_ex}. Similar to the case of WCE, $x_1$ is a staircase function of $x_0$, whereas $g_2$ is a scaled version of it to match the SNR constraint.

\section{Proposed Method}
The motivation for the DA algorithm is drawn the process of annealing in statistical physics, however, the method is founded on principles of information theory. Importantly, it replaces the stochastic operation of ``stochastic annealing" with the deterministic optimization of the effective expectation, namely, the free energy. DA introduces randomness into the optimization process, where the deterministic mappings (controllers) are replaced by random mappings. The optimization problem is recast as minimization of the expected cost subject to a constraint on the randomness (Shannon entropy) of the system. The resulting Lagrangian functional can be viewed as the free energy of a corresponding physical system whose Lagrange parameter is the ``temperature". The entropy constraint is gradually relaxed (by lowering the temperature) while the minimum cost is tracked, and deterministic mappings are obtained at the limit of zero entropy.

\subsection{Derivation}
Consider the structured mapping $g_1$ written as 
\begin{equation}
g_1(x_0)=g_{m_1}(x_0) \quad \text{for}\,\,\, x_0 \in \mathbb R_{m_1}
\end{equation}
where $m_1=\{1,2,...,M_{1,max}\}$. Each $g_{m_1}(x_0)$ is a parametric function called ``local model" and $\mathbb R_{m_1}$ denotes a partition region in input space. We have 
\begin{equation}
\bigcup_{m_1=1}^{M_{1,max}} \mathbb R_{m_1} = \mathbb R
\end{equation}

Effectively, the mapping $g_1$ is defined with a structure determined by two components: a space partition and a parametric local model per partition cell. While noting that local models can be in any parametric form, in this work we use affine local models given by 
\begin{equation}
g_{m_1}(x_0)=a_{m_1}x_0+b_{m_1}.
\end{equation}

We similarly define a structure for $g_2$:
\begin{equation}
g_2(x_0)=g_{m_2}(x_0) \quad \text{for}\,\,\, x_0 \in \mathbb R_{m_2}
\end{equation}
where $m_2=\{1,2,...,M_{2,max}\}$ and local models are affine: $g_{m_2}(x_0)=a_{m_2}x_0+b_{m_2}$. 

 The crucial idea in DA is to replace the deterministic partition of space by a random partition, i.e. to associate every input point with each one of regions \textit{in probability}. We define association probabilities 
\begin{equation}
p(m_i|x_0)  = \mathbb P\{x_0 \in \mathbb R_{m_i}\} = \mathbb P\{g(x_0) = g_{m_i}(x_0)\},
\label{eq:prob}
\end {equation} 
for $i=1,2$ and for all $m_i,x_0$. Let $M_1$ and $M_2$ denote the random variables representing the index of the local models. The system has a joint Shannon entropy which can be expressed as 
\begin{equation}
H(x_0,M_1,M_2)=H(x_0)+H(M_1|x_0) + H(M_2|x_0).
\end{equation}
since, by construction, $M_1$ and $M_2$ are independent given $x_0$. Since the first term is a constant determined by source, we conveniently remove it and define 
\begin{align}
H \triangleq \sum_{i=1,2} H(M_i|x_0) = - \sum_{i=1,2} \mathbb E\{\log p(M_i|x_0)\}
 \label{eq:entropy}
 \end{align} 
where $H$ is the average level of uncertainty in the partition of space.

In DA, the cost defined in (\ref{eq:cost1}) is minimized at prescribed levels of uncertainty as defined in (\ref{eq:entropy}). Accordingly, we construct the Lagrangian
\begin{equation}
F=D-TH,
\label{eq:free}
\end{equation}
as the objective function to be minimized, with $T$ being the Lagrange multiplier associated with the entropy constraint. The Lagrangian in (\ref{eq:free}) is referred to as the (Helmholtz) ``free energy", and the Lagrange parameter $T$ is called ``temperature", to emphasize the intuitively compelling analogy to statistical physics.

\subsection{Algorithm Sketch}
We begin by optimizing the free energy in (\ref{eq:free}) at high temperature, which effectively maximizes the entropy. Accordingly, the association probabilities are uniform and all local models are identical, in other words, there is effectively a single distinct local model. Thus, at the beginning we obtain the optimum solution when both $g_1$ and $g_2$ are restricted to be linear. As the temperature is decreased, a bifurcation point is reached where the current solution is no longer a minimum but a saddle point, such that there exist a better solution with the local models divided into two or more groups. As the current solution becomes a saddle point, a slight perturbation of local models will trigger the discovery of the new solution with increased number of effective local models. Such bifurcations are referred to as ``phase transitions", in the sense of symmetry breaking with increase in effective model size, and the corresponding temperatures are called ``critical temperatures". At the limit $T \rightarrow 0$, minimizing $F$ corresponds to minimizing $D$ directly, which produce deterministic mappings, as it is always advantageous to fully assign a source point to the model that makes the smallest contribution to $D$. 

Therefore, the practical algorithm consists of minimizing $F$, starting at a high value of $T$ and tracking the minimum while gradually lowering $T$. A brief sketch of the algorithm can be given as follows.
\begin{enumerate}
\item Start at high temperature, single model.
\item Duplicate local models.
\item Minimization of $F$.
\begin{enumerate}
\item Optimize $p(m_i|x_0)$ for all $m_i,x_0$, $i=1,2$.
\item Optimize $a_{m_i}$ and $b_{m_i}$, for all $m_i$, $i=1,2$, using gradient descent.
\item Optimize $W(\cdot,\cdot)$. 
\item Convergence test: If not converged go to (a).
\end{enumerate}
\item If temperature is above stopping threshold, lower temperature and go to step 2.
\end{enumerate}

\subsection{Update Equations}
Here we give the expressions for optimal $p(m_i|x_0)$ which are, naturally, Gibbs distributions.
 \begin{align}
 p(m_1|x_0) = \frac{e^{-D(m_1,M_2,x_0)/T}}{ \sum\limits_{m_1}e^{-D(m_1,M_2,x_0)/T} } \quad \forall m_1, x_0 \nonumber \\
p(m_2|x_0) = \frac{e^{-D(M_1,m_2,x_0)/T}}{ \sum\limits_{m_2}e^{-D(M_1,m_2,x_0)/T} } \quad \forall m_2, x_0
 \label{optimum prob}
 \end{align}
where $D(m_1,M_2,x_0)$ and $D(M_1,m_2,x_0)$ is the cost of associating $x_0$ with local model $g_{m_1}$ and $g_{m_2}$, respectively:
\begin{align}
D(m_1,M_2,x_0) &= k_1^2 g_{m_1}^2 + \mathbb E \{x_2^2 | M_1=m_1, X_0=x_0\} \nonumber \\
D(M_1,m_2,x_0) &= k_2^2g_{m_2}^2 + \mathbb E \{x_2^2 | M_2=m_2, X_0=x_0\}
\end{align}

The optimal second controller, $W(y_1,z)$, can be expressed in closed from as 
\begin{equation}
W(y_1,z) = \mathbb E \{ x_0 | y_1, z\}
\end{equation}
which can be written in terms of known quantities using the approach in \cite{emrah_dcc10}.

\section{Experimental Results}
The integrals in the algorithm are numerically calculated by sampling the space on the uniform grid, and the support of the Gaussian distribution is bounded to (5$\sigma$ to 5$\sigma$) interval.

\subsection{Original WCE}
\begin{figure}
\centering \includegraphics[width=1 \linewidth]{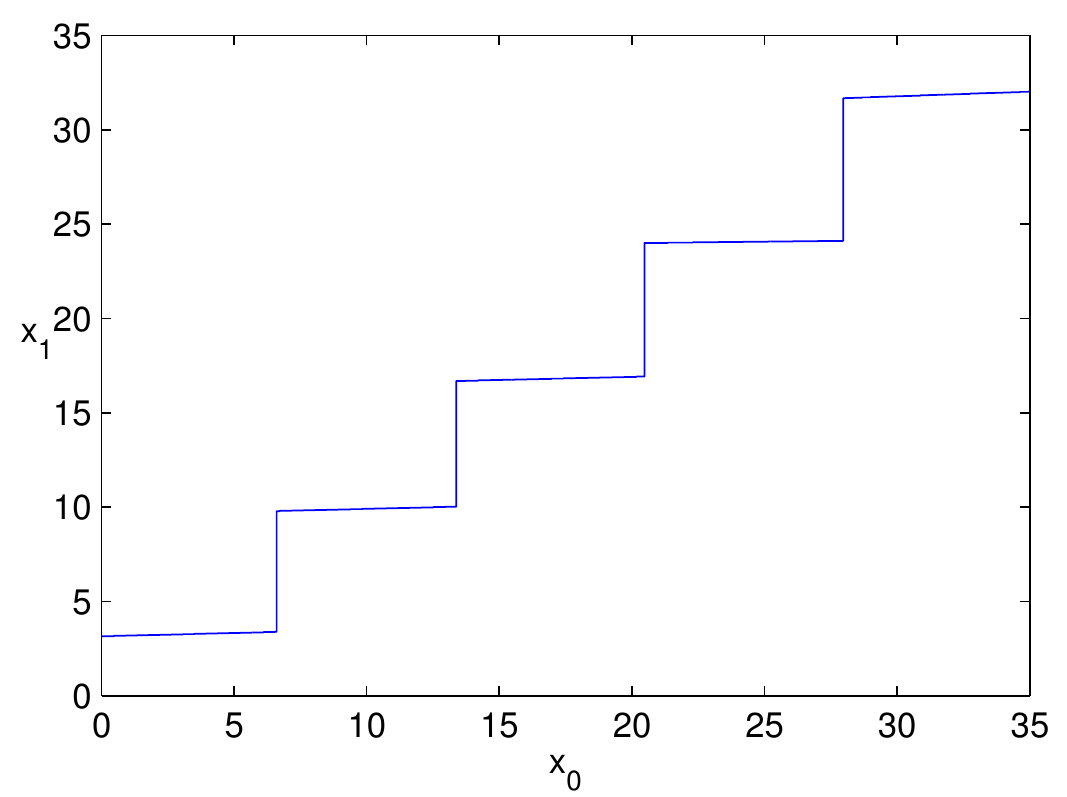}
\caption{The 5-step solution for the original Witsenhausen counterexample. Only positive half is shown.}
\label{original_mapping} 
\end{figure}

The DA method was applied to the original WCE problem and the results are reported in \cite{mehmetoglu2014deterministic}, where we obtained the lowest known  cost thus far, 0.16692291. We reproduce the results in comparison to prior work in Table \ref{prior}. The 5-step mapping we obtained after a sequence of phase transitions is given in Figure \ref{original_mapping}.

\begin{table}
\normalsize
\centering
\caption{Major Results for WCE}
\label{prior}
\begin{tabular}{rl}
\hline
Solution & Cost \\
\hline
Optimal Affine Solution & 0.961852 \\
1-step, Witsenhausen \cite{witsenhausen} & 0.404253 \\
2-step,\cite{deng} & 0.190 \\
Sloped 2.5 - step, \cite{baglietto} & 0.1701 \\
Sloped 3.5 - step, \cite{lee} & 0.1673132 \\
Sloped 3.5 - step,  \cite{li}& 0.1670790 \\
Sloped 4 - step, \cite{karlsson}& 0.16692462 \\
Sloped 5 - step, \cite{mehmetoglu2014deterministic} & 0.16692291\\
\hline
\end{tabular}
\end{table}

\subsection{Extension to Side Channel Variation}
In the experiments, we used the standard benchmark parameters that were used for the original WCE, that is, $k_1=0.2$ and $\sigma_{X_0}=5$. We have varied $k_2$ to obtain results at different side channel SNR values. Following the convention in \cite{martins}, we use $b_{SNR}=\sigma_{g_2} / \sigma_{n_2}$.

In Table \ref{comp} we compare the cost of our solutions (denoted by $D^*$) to the ones given in \cite{martins} (denoted by $D^M$), and the best affine mappings. Significant cost reductions can be observed. The relative improvement over the solution of \cite{martins} is listed in the last column.

\begin{remark}
When $b_{SNR}=0$, the problem degenerates to WCE, thus the cost is 0.1669, the best known to date.
\end{remark}

\begin{table}
\normalsize
\centering
\caption{Cost Comparison Table for the Side Channel Variation}
\label{comp}
\begin{tabular}{ccccc}
\hline
$b_{SNR}$ & Affine Cost & $D^M$ (\cite{martins}) & $D^*$ & $(\!D^M\!-\!D^*\!)/D^M$ \\
\hline
0.00 & 0.9600 & 0.1853 & 0.1669 & 0.10\\ 
2.37 & 0.7365 & 0.1546 & 0.0945 & 0.39\\ 
2.70 & 0.6802 & 0.1472 & 0.0837 & 0.43\\ 
5.62 & 0.3627 & 0.0852 & 0.0357 & 0.58\\ 
7.00 & 0.2814 & 0.0662 & 0.0264 & 0.60\\ 
9.57 & 0.1856 & 0.0497 & 0.0136 & 0.73\\
\hline
\end{tabular}
\end{table}

\begin{figure*}
\begin{minipage}[b]{0.5\linewidth}
  \centering
  \centerline{\includegraphics[width=8.5cm]{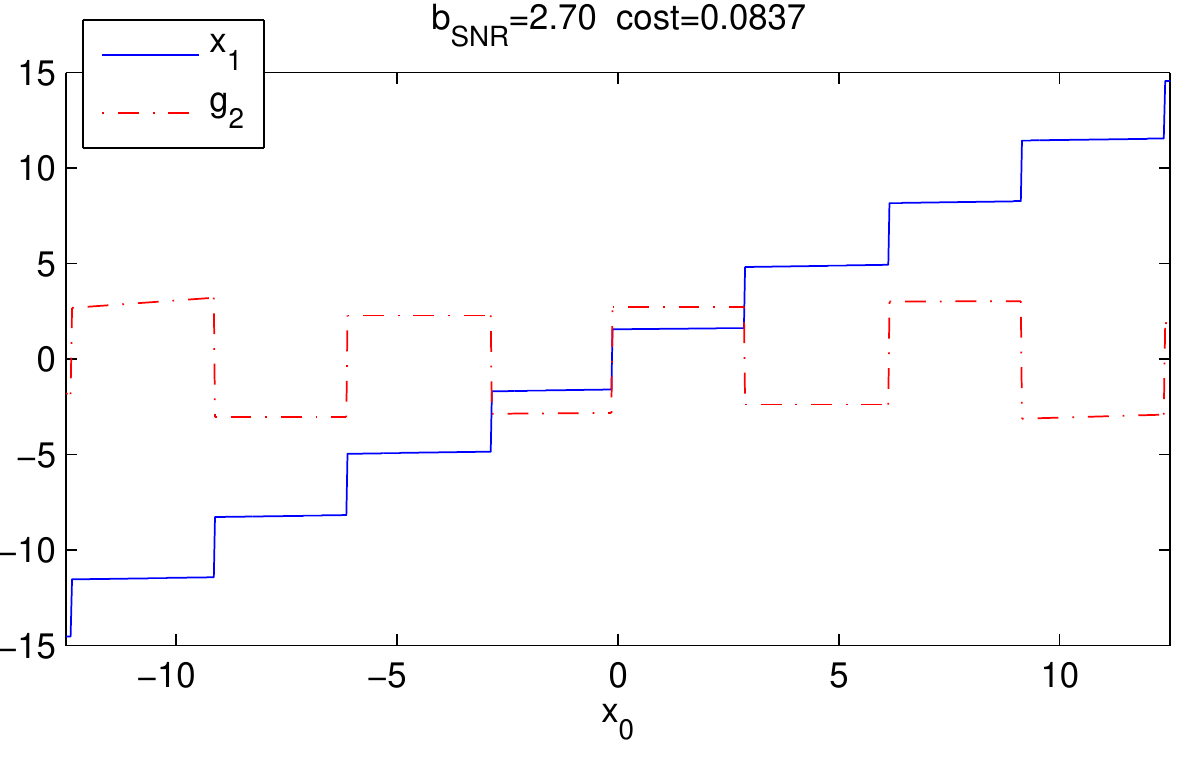}}
  \centerline{(a)}\medskip
\end{minipage}
\begin{minipage}[b]{0.5\linewidth}
  \centering
  \centerline{\includegraphics[width=8.5cm]{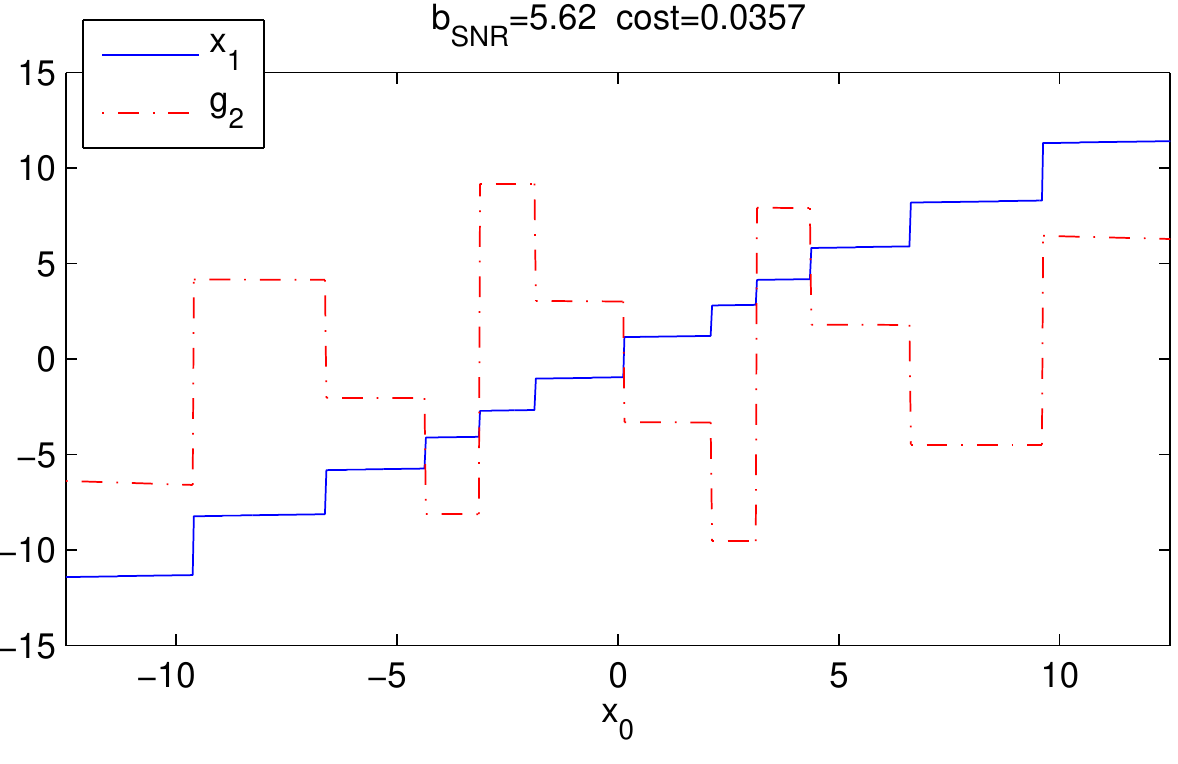}}
  \centerline{(b)}\medskip
\end{minipage}
\begin{minipage}[b]{0.5\linewidth}
  \centering
  \centerline{\includegraphics[width=8.5cm]{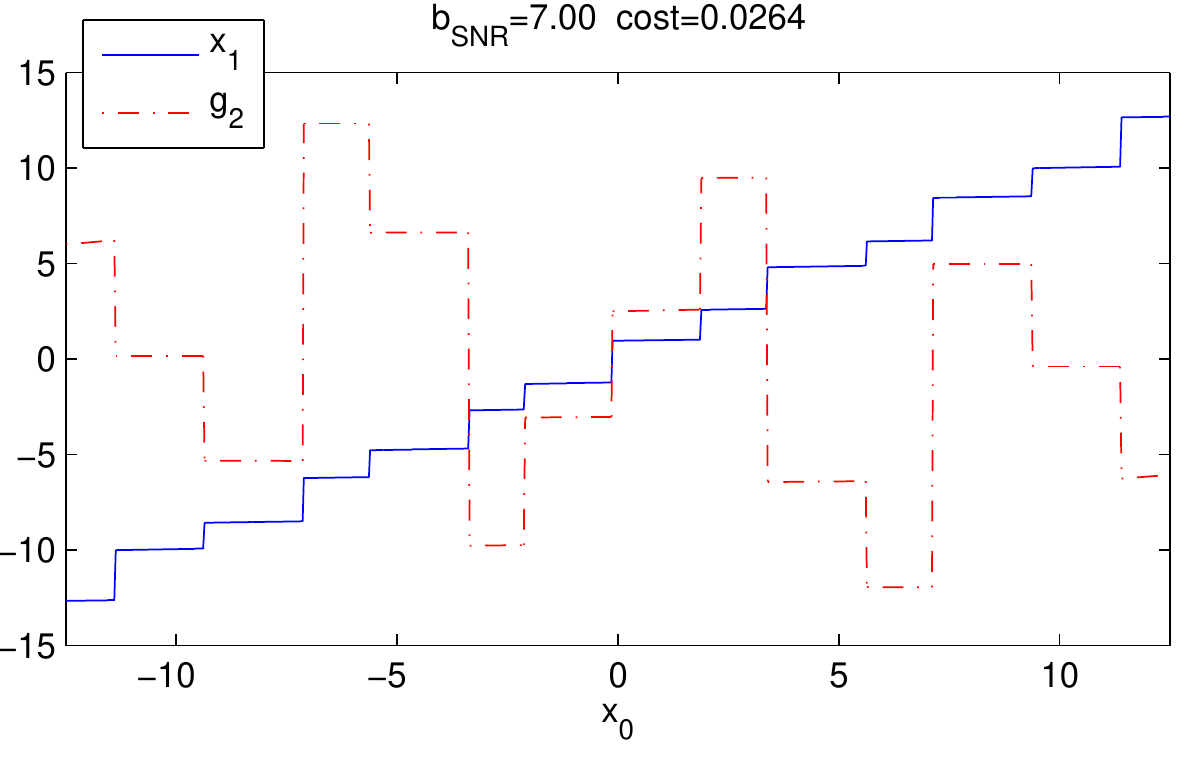}}
  \centerline{(c)}\medskip
\end{minipage}
\begin{minipage}[b]{0.5\linewidth}
  \centering
  \centerline{\includegraphics[width=8.5cm]{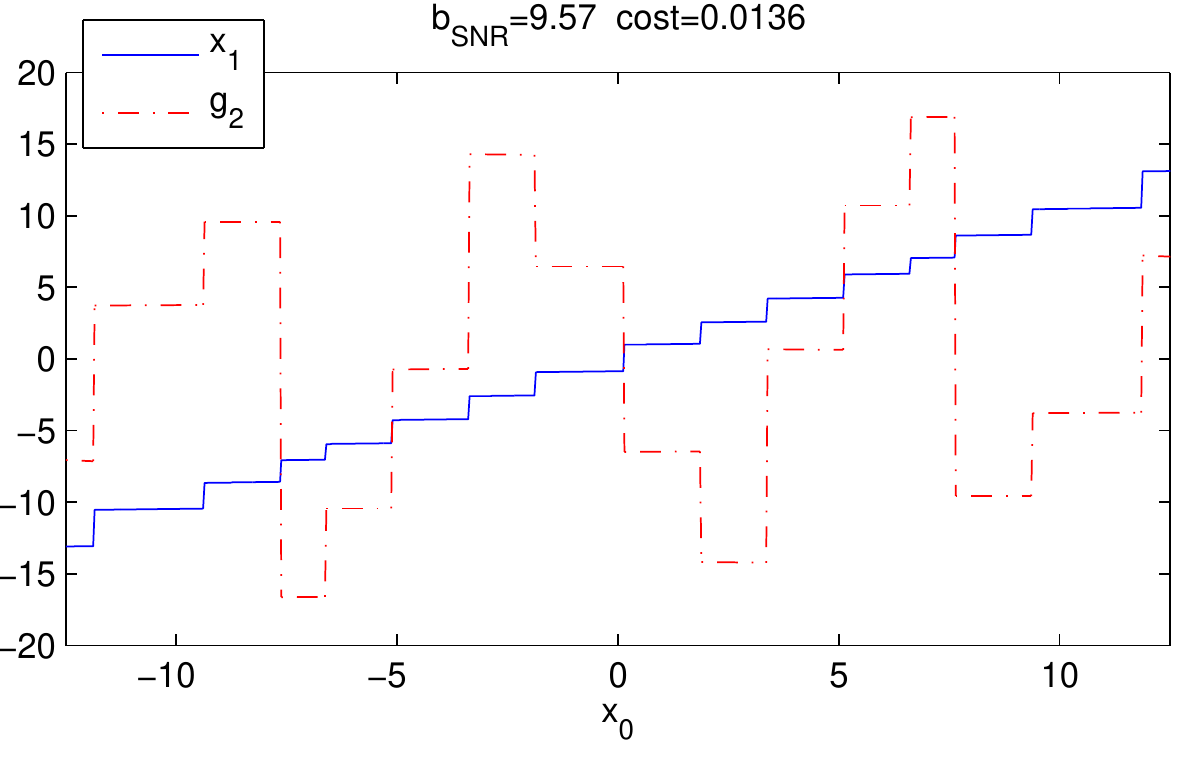}}
  \centerline{(d)}\medskip
\end{minipage}
\caption{Some of the mappings we obtained for the side channel variation problem. The first controller is plotted at various SNR levels.}
\label{examples}
\end{figure*}

We present several mappings obtained by our method in Figure \ref{examples}. Several interesting features of these mappings are observed. The mappings for $x_1$ are approximately staircase functions similar to the ones obtained for the original WCE problem, however, the steps get smaller and increase in number as the side channel SNR increases; that is, $x_1$ approaches $x_0$. Note that the control cost term in (\ref{cost}), $\mathbb E\{k_1^2g_1^2\}$, is minimum when $g_1=0$, in which case $x_1 = x_0$. This is, however, not optimum due to the estimation error at the second stage. Intuitively, as the second controller has access to better side information (i.e. at higher SNR), the estimation error is decreased and as observed in Figure \ref{examples}, $x_1$ tends to $x_0$. The relative improvement in cost, given in Table \ref{comp}, increases with SNR, which is consistent with the above observation. 

The mappings for the side channel, $g_2$, are highly irregular and the overall shape varies with SNR. This observation, together with the above for $x_1$, suggests that the mappings for $x_1$ and $g_2$ are not scale invariant. The discontinuities in $g_2$ and $x_1$ coincide as expected, as the discontinuities in side information $g_2$ signal those in $x_1$ to the second controller (estimator). 

{\bf Note}: Matlab code for our calculations of the total cost, including our decision functions can be found in \cite{witsen_webpage}. 

\section{Conclusions}
In this paper we extended our numerical method, introduced in prior work to obtain the best known solution for Witsenhausen's counterexample, to compute the elusive nonlinear mappings (controllers) in more involved decentralized control problems. As a test case we focused on the setting introduced in \cite{martins}, where it is motivated as a two stage noise cancellation problem. The mappings obtained are highly nontrivial and raise interesting questions about the functional properties of the optimal solution (mappings) in decentralized control, which are the focus of ongoing research.

\bibliographystyle{IEEEtran}

\bibliography{ref}

\begin{thebibliography}{10}
\providecommand{\url}[1]{#1}
\csname url@rmstyle\endcsname
\providecommand{\newblock}{\relax}
\providecommand{\bibinfo}[2]{#2}
\providecommand\BIBentrySTDinterwordspacing{\spaceskip=0pt\relax}
\providecommand\BIBentryALTinterwordstretchfactor{4}
\providecommand\BIBentryALTinterwordspacing{\spaceskip=\fontdimen2\font plus
\BIBentryALTinterwordstretchfactor\fontdimen3\font minus
  \fontdimen4\font\relax}
\providecommand\BIBforeignlanguage[2]{{%
\expandafter\ifx\csname l@#1\endcsname\relax
\typeout{** WARNING: IEEEtran.bst: No hyphenation pattern has been}%
\typeout{** loaded for the language `#1'. Using the pattern for}%
\typeout{** the default language instead.}%
\else
\language=\csname l@#1\endcsname
\fi
#2}}

\bibitem{mehmetoglu2014deterministic}
M.~Mehmetoglu, E.~Akyol, and K.~Rose, ``A deterministic annealing approach to
  {W}itsenhausen's counterexample,'' \emph{arXiv preprint arXiv:1402.0525,
  submitted to ISIT'14}, 2014.

\bibitem{witsenhausen}
H.~Witsenhausen, ``A counterexample in stochastic optimum control,'' \emph{SIAM
  Journal on Control}, vol.~6, no.~1, pp. 131--147, 1968.

\bibitem{basar2008variations}
T.~Ba\c{s}ar, ``Variations on the theme of the {W}itsenhausen counterexample,''
  in \emph{47th IEEE Conference on Decision and Control Proceedings
  (CDC)}.\hskip 1em plus 0.5em minus 0.4em\relax IEEE, 2008, pp. 1614--1619.

\bibitem{ubli}
C.~Choudhuri and U.~Mitra, ``On {W}itsenhausen's counterexample: {T}he
  asymptotic vector case,'' in \emph{Information Theory Workshop (ITW), 2012
  IEEE}, 2012, pp. 162--166.

\bibitem{martins}
N.~Martins, ``{W}itsenhausen's counter example holds in the presence of side
  information,'' in \emph{Decision and Control, 2006 45th IEEE Conference on},
  2006, pp. 1111--1116.

\bibitem{el2011network}
A.~El~Gamal and Y.~Kim, \emph{Network Information Theory}.\hskip 1em plus 0.5em
  minus 0.4em\relax Cambridge University Press, 2011.

\bibitem{wyner1975source}
A.~Wyner, ``On source coding with side information at the decoder,''
  \emph{Information Theory, IEEE Transactions on}, vol.~21, no.~3, pp.
  294--300, 1975.

\bibitem{ahlswede1975source}
R.~Ahlswede and J.~Korner, ``Source coding with side information and a converse
  for degraded broadcast channels,'' \emph{Information Theory, IEEE
  Transactions on}, vol.~21, no.~6, pp. 629--637, 1975.

\bibitem{rose1990statistical}
K.~Rose, E.~Gurewitz, and G.~Fox, ``Statistical mechanics and phase transitions
  in clustering,'' \emph{Physical review letters}, vol.~65, no.~8, pp.
  945--948, 1990.

\bibitem{rose1992vector}
------, ``Vector quantization by deterministic annealing,'' \emph{IEEE
  Transactions on Information Theory}, vol.~38, no.~4, pp. 1249--1257, 1992.

\bibitem{da}
K.~Rose, ``{Deterministic annealing for clustering, compression,
  classification, regression, and related optimization problems},''
  \emph{Proceedings of the IEEE}, vol.~86, no.~11, pp. 2210--2239, 1998.

\bibitem{emrah_dcc10}
E.~Akyol, K.~Rose, and T.~Ramstad, ``{Optimized analog mappings for distributed
  source channel coding},'' in \emph{Proceedings of IEEE Data Compression
  Conference}, 2010.

\bibitem{deng}
\BIBentryALTinterwordspacing
M.~Deng and Y.~Ho, ``An ordinal optimization approach to optimal control
  problems,'' \emph{Automatica}, vol.~35, no.~2, pp. 331 -- 338, 1999.
  [Online]. Available:
  \url{http://www.sciencedirect.com/science/article/pii/S0005109898001551}
\BIBentrySTDinterwordspacing

\bibitem{baglietto}
M.~Baglietto, T.~Parisini, and R.~Zoppoli, ``Numerical solutions to the
  {W}itsenhausen counterexample by approximating networks,'' \emph{Automatic
  Control, IEEE Transactions on}, vol.~46, no.~9, pp. 1471--1477, 2001.

\bibitem{lee}
J.~Lee, E.~Lau, and Y.-C. Ho, ``The {W}itsenhausen counterexample: a
  hierarchical search approach for nonconvex optimization problems,''
  \emph{Automatic Control, IEEE Transactions on}, vol.~46, no.~3, pp. 382--397,
  2001.

\bibitem{li}
N.~Li, J.~Marden, and J.~Shamma, ``Learning approaches to the {W}itsenhausen
  counterexample from a view of potential games,'' in \emph{Decision and
  Control, 2009 held jointly with the 2009 28th Chinese Control Conference.
  CDC/CCC 2009. Proceedings of the 48th IEEE Conference on}.\hskip 1em plus
  0.5em minus 0.4em\relax IEEE, 2009, pp. 157--162.

\bibitem{karlsson}
J.~Karlsson, A.~Gattami, T.~Oechtering, and M.~Skoglund, ``Iterative
  source-channel coding approach to {W}itsenhausen's counterexample,'' in
  \emph{American Control Conference (ACC), 2011}.\hskip 1em plus 0.5em minus
  0.4em\relax IEEE, 2011, pp. 5348--5353.

\bibitem{witsen_webpage}
\url{http://www.scl.ece.ucsb.edu/html/witsen.html}.

\end{thebibliography}

\end{document}